\documentclass[12pt]{iopart}
\usepackage{iopams} 
\usepackage{mathptmx}
\usepackage[utf8]{inputenc} 
\usepackage{graphicx}
\usepackage{epstopdf}
\usepackage[numbers,comma,square,sort&compress,merge]{natbib} 
\usepackage{upgreek} 
\usepackage{xspace}

\usepackage{url}

\usepackage{color,soul} 
\usepackage[colorinlistoftodos,prependcaption,textsize=footnotesize,textwidth=1.4cm]{todonotes}

\usepackage{hyperref}
\hypersetup{pdftitle={Near- and mid-infrared intersubband absorption in top-down GaN/AlN nano- and micropillars}, colorlinks, citecolor=blue}

\begin{document}
\graphicspath{{figures}}

\title{Near- and mid-infrared intersubband absorption in top-down GaN/AlN nano- and micropillars}
\author{Jonas L\"ahnemann$^{1,2}$, David A. Browne$^{1}$, Akhil Ajay$^{1}$, Mathieu Jeannin$^{3}$, Angela Vasanelli$^{3}$, Jean-Luc Thomassin$^{1}$, Edith Bellet-Amalric$^{1}$ and Eva Monroy$^{1}$}
\ead{laehnemann@pdi-berlin.de}
\address{$^1$ Universit\'e Grenoble-Alpes, CEA, INAC, PHELIQS, 17 av. des Martyrs, 38000 Grenoble, France}
\address{$^2$ Paul-Drude-Institut für Festkörperelektronik, Leibniz Institut im Forschugnsverbund Berlin e.V., Hausvogteiplatz 5--7, 10117 Berlin, Germany}
\address{$^3$ Universit\'e Paris Diderot, Sorbonne Paris Cité, Laboratoire Matériaux et Phénomènes Quantiques, UMR7162, 75013 Paris, France}


\begin{abstract}

We present a systematic study of top-down processed GaN/AlN heterostructures for intersubband optoelectronic applications. Samples containing quantum well superlattices that display either near- or mid-infrared intersubband absorption were etched into nano- and micropillar arrays in an inductively coupled plasma. We investigate the influence of this process on the structure and strain-state, on the interband emission and on the intersubband absorption. Notably, for pillar spacings significantly smaller ($\leq1/3$) than the intersubband wavelength, the magnitude of the intersubband absorption is not reduced even when 90\% of the material is etched away and a similar linewidth is obtained. The same holds for the interband emission. In contrast, for pillar spacings on the order of the intersubband absorption wavelength, the intersubband absorption is masked by refraction effects and photonic crystal modes. The presented results are a first step towards micro- and nanostructured group-III nitride devices relying on intersubband transitions.

\end{abstract}



\submitto{\NT}


\maketitle

\section{Introduction}

Engineering intersubband transitions between confined states in the conduction band of group-III arsenide quantum wells has led to the fabrication of both detectors and emitters across the mid- and far-infrared (IR) spectral ranges. In recent years, group-III nitride semiconductors have evolved into a promising alternative for the realization of intersubband technologies \cite{Lim_2017b}. Advantages of the group-III nitrides can be found in their large conduction band offset, e.g.\ 1.8~eV for GaN/AlN \cite{Binggeli_2001,Tchernycheva_2006,Edmunds_2013}, which offers the possibility to extend the operating range of intersubband devices into the near-IR spectral range, including the 1.3--1.55~$\upmu$m wavelength window. Devices for fiber-optic telecommunications could also profit from the sub-picosecond intersubband relaxation times \cite{Iizuka_2000,Gmachl_2001}. Furthermore, the large longitudinal-optical (LO) phonon energy (91~meV) in GaN opens prospects for room-temperature operation of devices designed for THz frequencies \cite{Bellotti_2008,Edmunds_2014,Lim_2017b}.

Besides planar quantum well structures, a variety of three-dimensional geometries have been proposed in the past years. In particular, nanowires and micropillars are explored as a way to independently control the electrical and the optical device cross section by changing the diameter-to-pitch ratio in the array. Furthermore, the three-dimensional carrier confinement in nanowire heterostructures opens new possibilities to tune the carrier relaxation time. Such structures can be achieved by bottom-up growth or top-down etching. On the one hand, near-IR intersubband absorption could be demonstrated for GaN/AlN superlattices grown on self-assembled GaN nanowires \cite{Beeler_2014}. In such nanowires, dispersed and processed into miniaturized photodetectors, an intersubband photocurrent was detected under near-IR illumination \cite{Lahnemann_2017a}. Despite the large lattice-mismatch between GaN and AlN, bottom-up nanowire heterostructures have low dislocation densities thanks to the elastic strain relaxation at the nanowire surface. However, this advantage comes at the expense of a wire-to-wire inhomogeneity of the superlattice dimensions inherent to the self-assembled growth \cite{Lahnemann_2016,Ajay_2017a}. As a consequence, the intersubband absorption is spectrally broad when compared to planar quantum wells \cite{Kandaswamy_2008} or quantum dots \cite{Vardi_2009}. On the other hand, nano- and micropillar arrays for THz emission have been achieved by lithographically defined top-down etching of (Al,Ga)As heterostructures \cite{Amanti_2013, Krall_2014, Krall_2015}. Particularly, Amanti \emph{et al.}~\cite{Amanti_2013} have shown THz quantum-cascade lasing in a micropillar array, which due to its sub-wavelength dimensions could be considered as photonic metamaterial. In the case of photodetectors, by optimizing the pillar dimensions and spacing based on electrodynamics simulations of the light incoupling at the absorbed wavelength, pillar detectors can be expected to outperform planar detectors, while at the same time lifting selection rules concerning the polarization of the incoming light \cite{Karimi_2018}.

The aim of the current study is to investigate the influence of the top-down etching of nano- and micropillar arrays on the intersubband absorption in GaN/AlN heterostructures, to assess the suitability of this geometry for group-III nitride intersubband devices. To this end, we grew two samples designed to present intersubband transitions in the near- and mid-IR spectral ranges, respectively, and these were processed into  micro-/nanopillars. Here, we show that when the spacing of the pillar array is comparable to the wavelength of the investigated spectral region, the intersubband absorption is masked by refraction effects and photonic crystal modes in the IR transmittance spectra. For subwavelength spacing, the intersubband absorption is recovered with similar linewidth as in the as-grown material. Furthermore, our work demonstrates that the amplitude of both the interband emission and the intersubband absorption are attenuated only weakly in comparison to the reduction of material during the top-down processing.

\section{Experiments and Methods}

\begin{figure}
\centering
\includegraphics*[width=10cm]{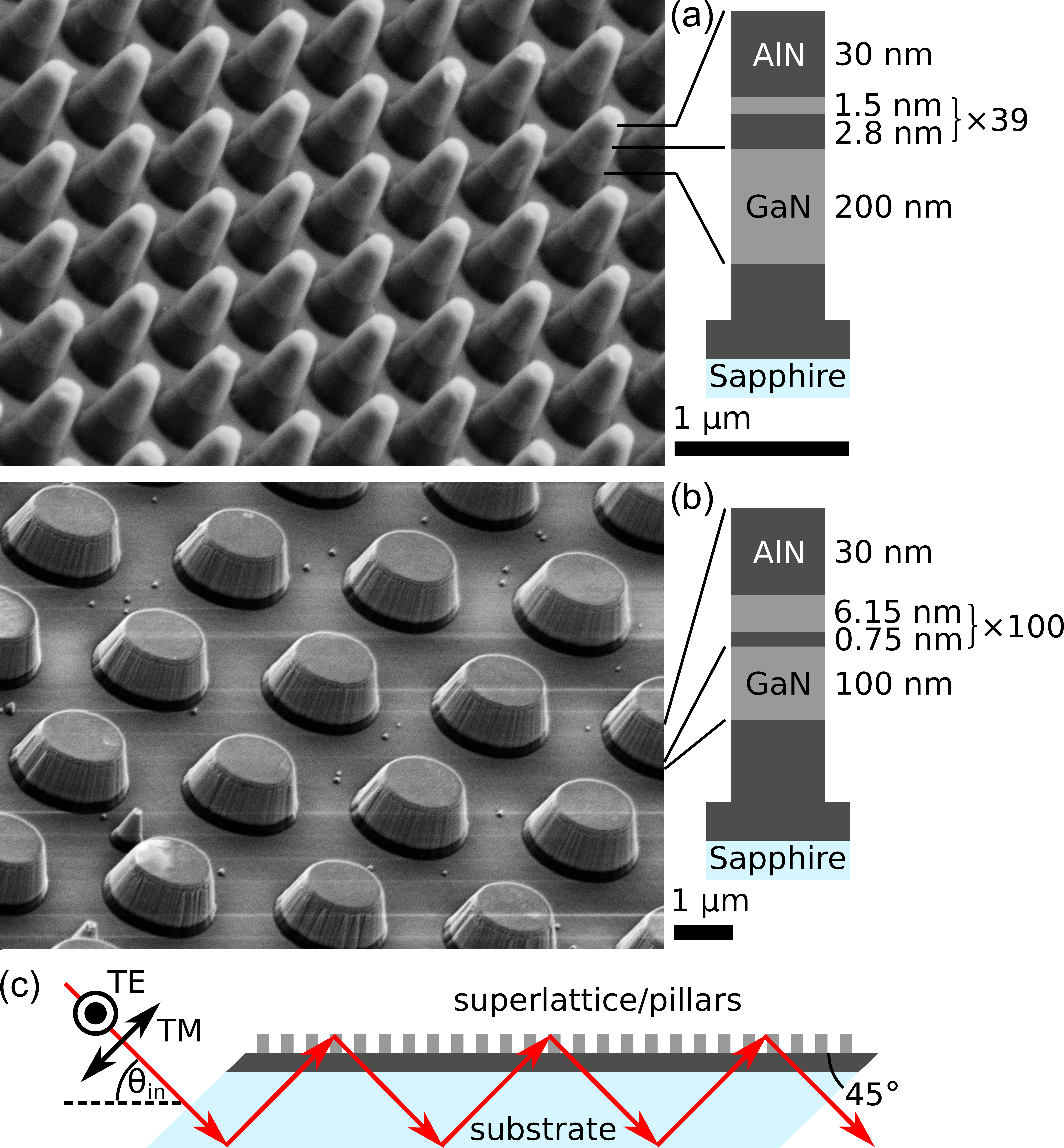}
\caption{\label{fig:sem}Bird's-eye view SEM images and sketches of the layer stacks for samples (a) NIR etched into pillars with $d=0.2~\upmu$m and (b) MIR with $d=1.5~\upmu$m. (c) Sketch of the waveguide geometry used for the FTIR measurements, indicating the incident angle of the light on the side facet $\theta_\mathrm{in}$ and the criterion to define TE and TM polarization.}
\end{figure}

The investigated samples consist of superlattices of highly doped GaN/AlN quantum wells (QWs) grown by plasma-assisted molecular-beam epitaxy (PAMBE)  on a 1~$\upmu$m thick AlN-on-sapphire template. The growth conditions were described elsewhere \cite{Kandaswamy_2008}. Prior to the growth of the superlattice, a 200~nm thick undoped GaN layer was deposited under Ga-rich conditions. This GaN buffer remains partially strained on AlN \cite{Bellet-Amalric_2004}, which allows the growth of the superlattice without cracking. After the growth of the superlattice, a 30~nm thick AlN cap layer was deposited to prevent the depletion of the QWs due to surface states. The study was performed on two samples. The first of them (sample NIR) consists of 39 periods of 1.5~nm thick GaN:Si QWs separated by 2.8~nm thick AlN barriers. The second sample (sample MIR) consists of 100 periods of 6.15~nm thick GaN:Si QWs separated by 0.75~nm thick AlN barriers. Silicon was incorporated in the GaN QWs to reach a surface doping density of $1.6~\times~10^{13}$~cm$^{-2}$. The electronic levels in the QWs were calculated using the \textbf{k}$\cdot$\textbf{p} module of the self-consistent Schrödinger-Poisson solver Nextnano \cite{Birner_2007} with the material parameters described in Ref.~\citenum{Kandaswamy_2008}. The difference in energy between ground and first excited state of the electron in the QWs (e$_2$--e$_1$), including corrections for many-body effects \cite{Edmunds_2013,Beeler_2014}, was calculated to be 0.81~eV ($\lambda\approx 1.53~\upmu$m) for sample~NIR and 0.29~eV ($\lambda\approx 4.33~\upmu$m) for sample MIR, respectively.

Micro- and nanopillars were etched using a top-down process. To this end, a Ti/Ni (5/50~nm) etch mask of circular structures on a hexagonal grid was defined by lift-off. The pattern was written by laser lithography for pillars with a diameter $d=1.5~\upmu$m (both samples) and electron beam lithography for diameters $d=0.4$, 0.2 and 0.1 $\upmu$m (only sample NIR). The pillar dimensions and pitch (center-to-center distance) are summarized in table~\ref{tab:samples}. The material around the pillars was removed with an inductively-coupled plasma (ICP) etch. The following parameters, optimized for steep sidewalls \cite{Albert_2014}, were used: 10~sccm Cl$_2$ and 25~sccm Ar with an ICP power of 300~W and a radio-frequency power of 150~W at room temperature at a pressure of 5 mTorr. A 3~min etch leads to an etch depth of approximately 1100~nm. Subsequently, the metal mask was removed with FeCl$_3$ and HF. Exemplary scanning electron microscopy (SEM) images of the resulting nano- and micropillars are shown in figure~\ref{fig:sem}(a) for sample NIR with $d=0.2~\upmu$m and in figure~\ref{fig:sem}(b) for sample MIR with $d=1.5~\upmu$m together with sketches of the respective superlattice structures.

High-resolution x-ray diffraction (XRD) of the as-grown and processed samples was performed using a Rigaku SmartLab X-Ray diffractometer system with a 2-bounce Ge(220) monochromator and a 0.228$^\circ$ long plate collimator in front of the detector. The influence of etching on the interband emission was investigated by photoluminescence (PL) spectroscopy under excitation with a continuous-wave solid-state laser (wavelength $\lambda=244$~nm, laser power of $0.4~\upmu$W) focused to a spot with a diameter of $\approx100~\upmu$m. The sample emission is passed through a grating monochromator and is detected using an ultraviolet-enhanced charge-coupled device. PL measurements were carried out at 300 and 10~K. The intersubband absorption was measured at room temperature in a Bruker 70v Fourier transform infrared (FTIR) spectrometer. For this purpose, the samples were polished into a waveguide geometry with the side facets for in- and out-coupling polished at 45$^\circ$ as sketched in figure~\ref{fig:sem}(c). Thereby, the sample can be rotated to change the incident angle. The spectrometer is equipped with a halogen lamp (near-IR), an Hg discharge lamp (mid-IR), a CaF$_2$ beam splitter, and a nitrogen-cooled mercury-cadmium-telluride detector. The transverse electric (TE) and transverse magnetic (TM) polarization components were distinguished by polarizing the incident light with a linear polarizer placed in front of the sample. As the selection rules restrict the intersubband absorption to TM polarized light, the TM transmittance spectra were divided through the transmittance in TE polarization to correct for the system response. 

The near-IR transmittance and absorption spectra of the pillar arrays were simulated using a commercial finite element solver (Comsol v5.3a). As an exact modeling of the photonic crystal structure lies beyond the scope of this paper, the simulations were restricted to a vertical cross-section in the plane of light propagation. The correct coupling geometry, the semiconductor layer sequence, the etch depth with a tapering angle of 8$^\circ$ [c.f.\ figure~\ref{fig:sem}] and the correct pitch were taken into account in the definition of the structure, which was thus modelled as a periodic ridge grating. Even if it remains an  approximation, such a simple model already describes the main features of the experimental spectra. The transmittance was calculated by exciting the pillar array with a plane wave from the substrate side and integrating the field reflected towards the substrate. The intersubband absorption in the QWs was calculated by integrating the resistive losses in the QWs volume. The anisotropic refractive index of the QWs was determined using a semi-classical treatment. Starting from the electronic band structure and accounting for non-parabolicity, the different matrix elements associated to all the intersubband transitions were calculated, explicitly taking into account the doping level in the QWs and the depolarization effect \cite{Warburton_1996}. 

\section{Results}

\subsection{Structure and strain-state}

\begin{table*}{}
\caption{\label{tab:samples}: Summary of the sample characteristics, including pillar diameters and pitches. Structural parameters determined from XRD measurements include the superlattice (SL) period, the FWHM ($\Delta\omega$) of the $\omega$-scans across the (0002) reflections of the SL, the AlN substrate and the GaN buffer, as well as the GaN $c$ lattice constant in the buffer layer and the strain in the buffer layer $\epsilon_{zz}$. From the PL measurements, the low-temperature integrated intensity $I_\mathrm{PL}(10~\mathrm{K})$ normalized to its value for the as-grown sample, as well as the ratio of room- and low- temperature intensities $I_\mathrm{PL}(300~\mathrm{K})/I_\mathrm{PL}(10~\mathrm{K})$ are given. Regarding the FTIR measurements, the center $\lambda$ and FWHM of the absorption band, measured before and after etching, are given for the different samples.}\vspace{1mm}
{\footnotesize
\begin{tabular}{lcccccccc}
\hline\hline
Sample & NIR & & & & & \hspace{2mm} & MIR &\\
\hline
Diameter ($\upmu$m) &  & 1.5 & 0.4 & 0.2 & 0.1 & & & 1.5 \\
Pitch ($\upmu$m) & as-grown & 3.0 & 1.2 & 0.6 & 0.3 & & as-grown & 3.0 \\
\hline
SL period ($\pm0.1$~nm) & 4.3 & 4.0 & 4.2 & 4.2 & 4.4 & & 7.1 & 6.9 \\
$\Delta\omega_\mathrm{SL}$ ($^\circ$) & 0.18 & 0.38 & 0.44 & 0.28 & 0.44 & & 0.21 & 0.27 \\
$\Delta\omega_\mathrm{AlN}$ ($^\circ$) & 0.06 & 0.06 & 0.06 & 0.05 & 0.05 & & 0.06 & 0.07 \\
$\Delta\omega_\mathrm{GaN}$ ($^\circ$) & 0.16 & 0.42 & 0.33 & 0.30 & 0.43 & & 0.33 & 0.36 \\
$c_\mathrm{GaN}$ ($\pm0.002$~\AA) & 5.208 & 5.192 & 5.185 & 5.183 & 5.186 & & 5.211 & 5.198 \\
$\epsilon_{zz}$ ($\pm4\times10^{-4}$) & $4.6\times10^{-3}$ & $1.5\times10^{-3}$ & $3\times10^{-4}$ & $-3\times10^{-4}$ & $4\times10^{-4}$ & & $5.2\times10^{-3}$ & $2.7\times10^{-3}$ \\
\hline
$I_\mathrm{PL}(10~\mathrm{K})$ (normalized) & 1.0 & 0.68 & 0.73 & 0.96 & 0.65 & & 1 & 0.75\\
$I_\mathrm{PL}(300~\mathrm{K})/I_\mathrm{PL}(10~\mathrm{K})$ & 0.10 & 0.08 & 0.07 & 0.03 & 0.05 & & -- & --\\
\hline
$\lambda_\mathrm{FTIR}(\mathrm{as~grown})$ ($\upmu$m) & & -- & 1.43 & 1.54 & 1.51 & & & 4.4\\
$\lambda_\mathrm{FTIR}(\mathrm{etched})$ ($\upmu$m) & & -- & 1.38 & 1.42 & 1.47 & & & 4.6\\
\multicolumn{2}{l}{$\mathrm{FWHM}_\mathrm{FTIR}(\mathrm{as~grown})$ ($\upmu$m)} & -- & 0.28 & 0.33 & 0.32 & & & 1.4 \\
\multicolumn{2}{l}{$\mathrm{FWHM}_\mathrm{FTIR}(\mathrm{etched})$ ($\upmu$m)} & -- & 0.28 & 0.31 & 0.4 & & & 1.8 \\
\hline \hline
\end{tabular}
}
\end{table*}

\begin{figure}
\centering
\includegraphics*[width=10cm]{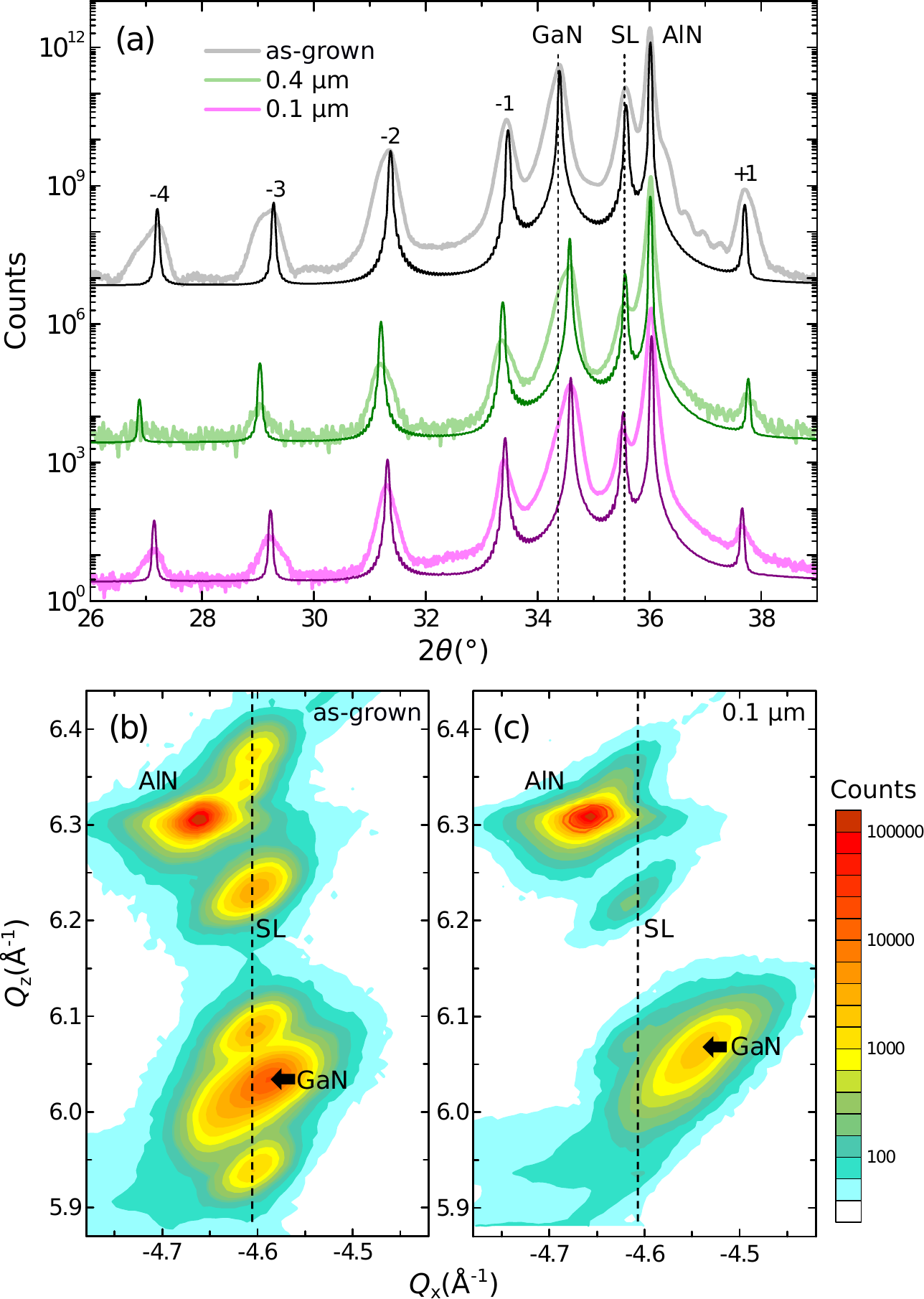}
\caption{\label{fig:xrd}(a) XRD $\omega-2\theta$ scans around the $(0002)$ reflection of sample NIR as-grown and etched into pillars with $d=0.4$ and 0.1~$\upmu$m. Superimposed are the simulated profiles. Additionally, reciprocal space maps around the $(\bar{2}025)$ reflection are shown for sample NIR (b) as-grown and (c) etched into nanopillars with $d=0.1$~$\upmu$m.}
\end{figure}

The structural characteristics of the GaN/AlN superlattice and the influence thereon of the top-down processing were investigated by high-resolution XRD. Figure~\ref{fig:xrd}(a) shows exemplary $\omega-2\theta$ profiles around the $(0002)$ reflection for sample NIR as-grown and etched into nanopillar arrays. Theoretical calculations using the Rigaku SmartLab Studio software are superimposed upon the experimental data. The superlattice period extracted from these $\omega-2\theta$ scans is summarized in table~\ref{tab:samples} for all investigated samples, together with the full width at half maximum (FWHM) of the $\omega$-scan around the $(0002)$ reflection of the superlattice ($\Delta\omega_\mathrm{SL}$), the AlN template ($\Delta\omega_\mathrm{AlN}$), and the GaN buffer layer ($\Delta\omega_\mathrm{GaN}$). Variations in the superlattice period are due to a gradient of the growth rate along the wafer resulting from the radial variation of the active nitrogen flux impinging on the substrate. In the diffractogram of the as-grown material, the angular location of the reflection from the GaN buffer layer confirms that it remains partially strained by the AlN substrate. To be precise, the as-grown out-of-plane deformation is $\epsilon_{zz}=(c-c_{0})/c_{0}=(4.6\pm0.4)\times10^{-3}$, with $c$ being the measured out-of-plane lattice constant and $c_0=5.1850$~\AA{} being the lattice constant for relaxed GaN. This value corresponds to a relaxation of GaN on AlN by only 65\%, with the relaxation defined as $R(\%)=100 \times (c_\mathrm{FS}-c)/(c_\mathrm{FS}-c_{0})$ where $c_\mathrm{FS}=5.2517$~\AA{} is the out-of-plane lattice parameter of GaN fully strained on AlN.

Regarding the FWHM of the $\omega$-scans around the $(0002)$ reflections of the different layers ($\Delta\omega$ values in table~\ref{tab:samples}), the  increase of $\Delta\omega_\mathrm{GaN}$ and $\Delta\omega_\mathrm{SL}$ in the etched samples in comparison with the as-grown structures is explained, on the one hand, by the nanostructuration, and, on the other hand, by the expected strain relaxation towards the pillar side walls. The $\omega-2\theta$ reflections assigned to the superlattice do not present any significant angular shift before and after etching. However, the intensity of the peaks related to the superlattice and to the GaN buffer layer decreases to about 10\% of the as-grown value, when taking the intensity of the AlN substrate as a reference. This decrease is explained by the fact that there is only 10\% of the material remaining after the etching process. In contrast, the reflection associated to GaN exhibits a clear, systematic shift towards larger angles in the etched structures, which is an indication of strain relaxation. The lattice parameter $c_\mathrm{GaN}$ extracted from these data and the strain along the growth axis, $\epsilon_{zz}$, are listed in table~\ref{tab:samples}. Full relaxation (within the error bar of the measurements of $\pm 4$\%) of the GaN buffer is obtained for pillars with a diameter $\leq0.4~\upmu$m. Previous studies on GaN nanopillars have seen full relaxation only at the point where the pillar length (here GaN segment length) exceeds the pillar diameter \cite{Hugues_2013}.

The determination of the strain state of the superlattice requires information from asymmetric reflections. Therefore, reciprocal space maps around the $(\bar{2}025)$ reflection were collected for selected samples. Figures~\ref{fig:xrd}(b) and (c) show the reciprocal space maps for the sample NIR as-grown and etched to nanopillars with a diameter of $d=0.1~\upmu$m, respectively. In both cases, several satellites of the superlattice reflection can be identified, which appear at an intermediate $Q_\mathrm{x}$ vector between the AlN substrate and the GaN buffer layer. The $Q_\mathrm{x}$ location of the reflections related to the superlattice does not vary with etching; it is only the reflection from the GaN buffer layer that exhibits a drastic shift, consistent with the observations in the $\omega-2\theta$ scans. This behavior indicates that the strain state of the superlattice is decoupled from the strain state of the underlying GaN layer. Extracting the average lattice parameter $c$ of the superlattice from the $\omega-2\theta$ diffractogram and its average in-plane lattice parameter from the reciprocal space map, we estimate an average Al concentration in the structure of $65\pm1$\%, and that the superlattice is fully relaxed. This result explains the insensitivity of the superlattice to the strain state of the GaN underlayer and is in agreement with Kandaswamy \emph{et al.} \cite{Kandaswamy_2009a}, who have shown that full relaxation of similar GaN/AlN superlattices can be achieved after the growth of about 10 GaN/AlN periods independently of the substrate. The simulations of the $\omega-2\theta$ scans in Figure~\ref{fig:xrd}(a) take into account the composition and strain states extracted from the reciprocal space maps.

\begin{figure}
\centering
\includegraphics*[width=12cm]{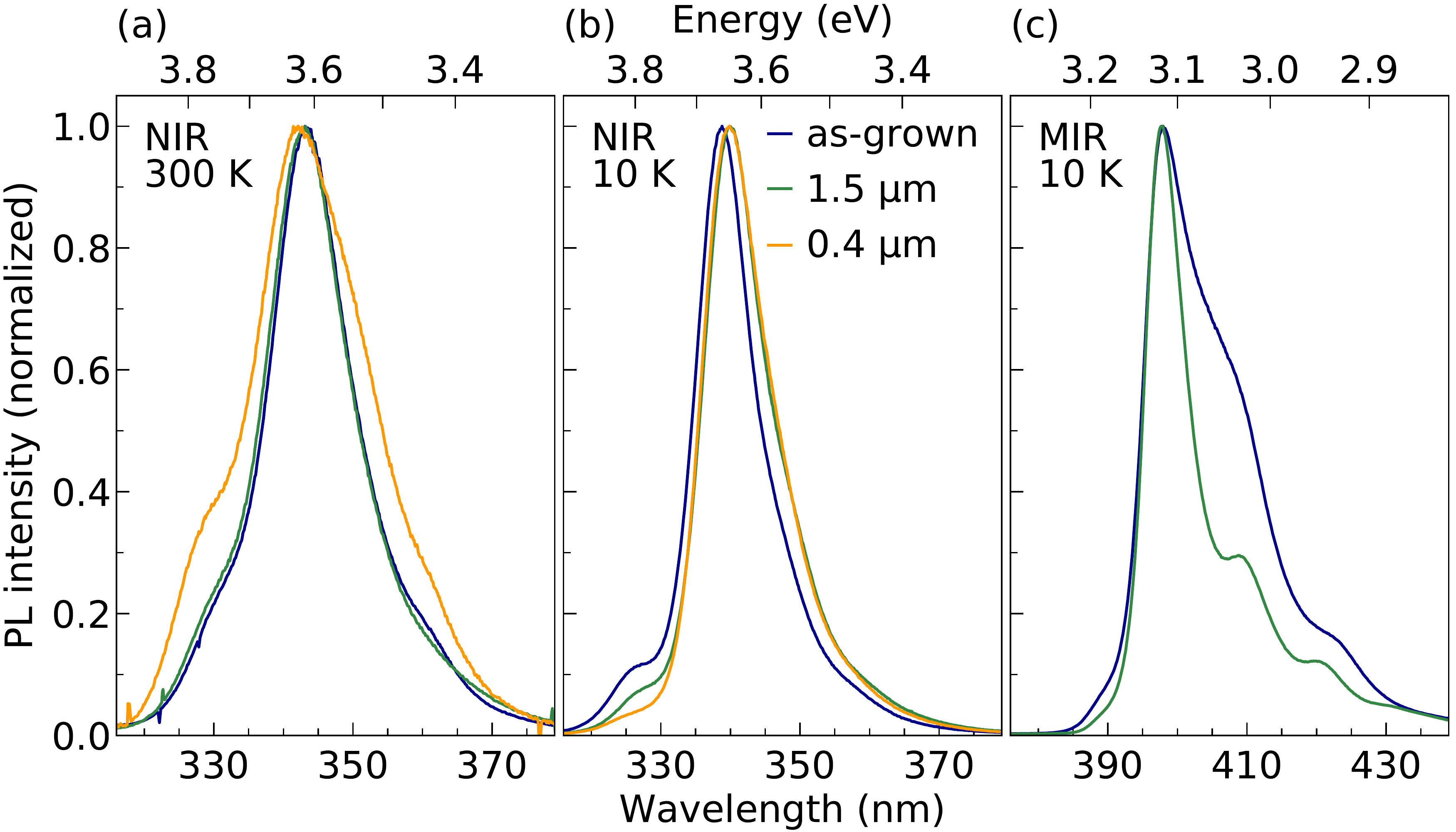}
\caption{\label{fig:pl}(a) Room-temperature and (b) low-temperature PL spectra for sample NIR in the as-grown state, as well as for pillar diameters of 1.5 and 0.4~$\upmu$m. (c) Low-temperature PL spectra for sample MIR as-grown and etched into 1.5~$\upmu$m pillars.}
\end{figure}

\subsection{Interband emission}

To get a first idea about the impact of the top-down processing on the optical characteristics, we have studied the PL emission of the heterostructures. Exemplary spectra from sample NIR recorded at room temperature (300~K), and from both samples NIR and MIR at low temperature (10~K) are presented in Figs.~\ref{fig:pl}(a)--(c). For sample NIR, the emission is centered around 343~nm at room temperature and around 339~nm at 10~K. In comparison, \textbf{k}$\cdot$\textbf{p} calculations predict interband transition wavelengths of 342~nm at 300~K and 335~nm at 10~K, which confirms the validity of these simulations (note that excitonic effects are not taken into account). For sample MIR, the low-temperature PL emission wavelength of 400~nm also coincides with the calculated interband transition energy. However, for this sample, the PL signal could not be detected at 300~K, a consequence of the strong quantum confined Stark effect in these wide QWs. Note that both samples do not exhibit any significant shift of the PL emission wavelength after etching. This behavior is in agreement with the XRD results for the superlattice, which is relaxed already after growth and shows no variation of the strain state after etching.

The measured low temperature PL intensities, normalized to the emission of the as-grown samples, and the ratios between the integrated intensities $I_\mathrm{PL}$ at 300 and 10~K (only for sample NIR) are summarized in table~\ref{tab:samples}. Strikingly, the emission intensity at 10~K is not significantly degraded by the etching, the attenuation amounts to less than a factor of 2. Considering that the top-down processing removes about 80\% of the material in the case of the micropillars ($\mathrm{pitch}=2\times\mathrm{diameter}$) and 90\% of the material in the case of the nanopillars ($\mathrm{pitch}=3\times\mathrm{diameter}$), this result indicates that the in- and outcoupling of the ultraviolet light is significantly improved for the pillar arrays as compared to the planar layers \cite{Reddy_2016,Hauswald_2017}. At the same time, compared with the as-grown planar reference, the ratio $I_\mathrm{PL}(300~\mathrm{K})/I_\mathrm{PL}(10~\mathrm{K})$ shows a trend to smaller values with decreasing feature size, with the reduction being limited to a factor of 3 in the case of the 0.2~$\upmu$m sample. Apparently, the processing induces some nonradiative defects at or close to the surface, which are activated at room temperature. However, it is a notable result -- also important for intersubband device applications -- that, even for the smallest pillar diameter, the non-radiative recombination is only moderately enhanced. Note that no additional processing \cite{Li_2016a} for defect removal or surface passivation has been applied after the ICP etching process.

\begin{figure*}
\centering
\includegraphics*[width=16cm]{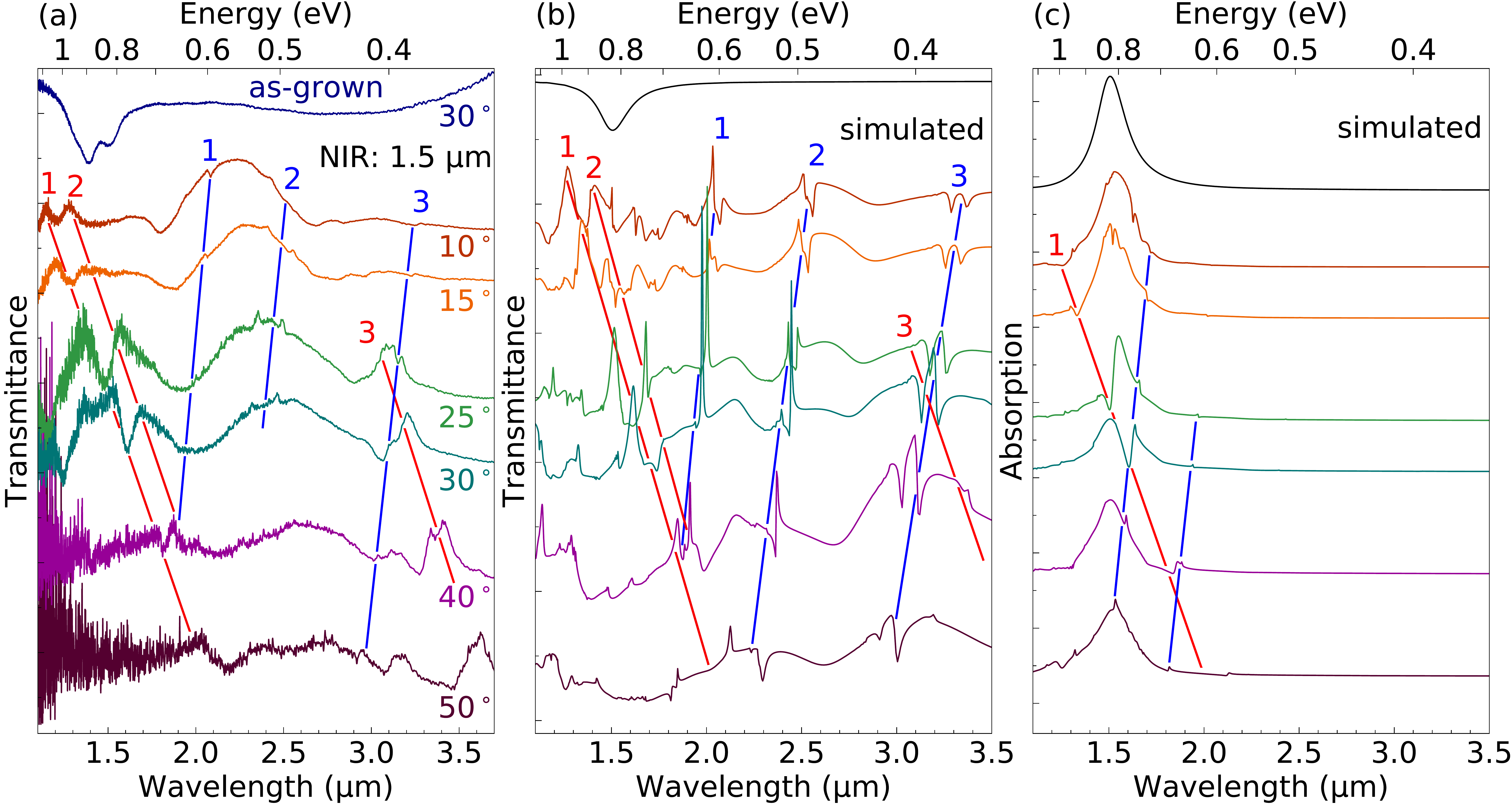}
\caption{\label{fig:ftir1}(a) Measured FTIR transmittance of the top-down-processed sample NIR ($d=1.5~\upmu$m) as a function of the incident angle $\theta_\mathrm{in}$, as defined in figure~\ref{fig:sem}(c). For clarity, the spectra are shifted vertically. The transmittance of the as grown sample is presented on top for comparison. (b) Simulated transmittance and (c) absorption of the pillar array. Both in the measurement and simulations, the TM-polarized transmittance is corrected by dividing through the TE-polarized transmittance. The red and blue lines are guides for the eye marking features that, with increasing incident angle, shift to the red and blue, respectively.}
\end{figure*}

\subsection{Intersubband absorption}

\begin{figure*}
\centering
\includegraphics*[width=13cm]{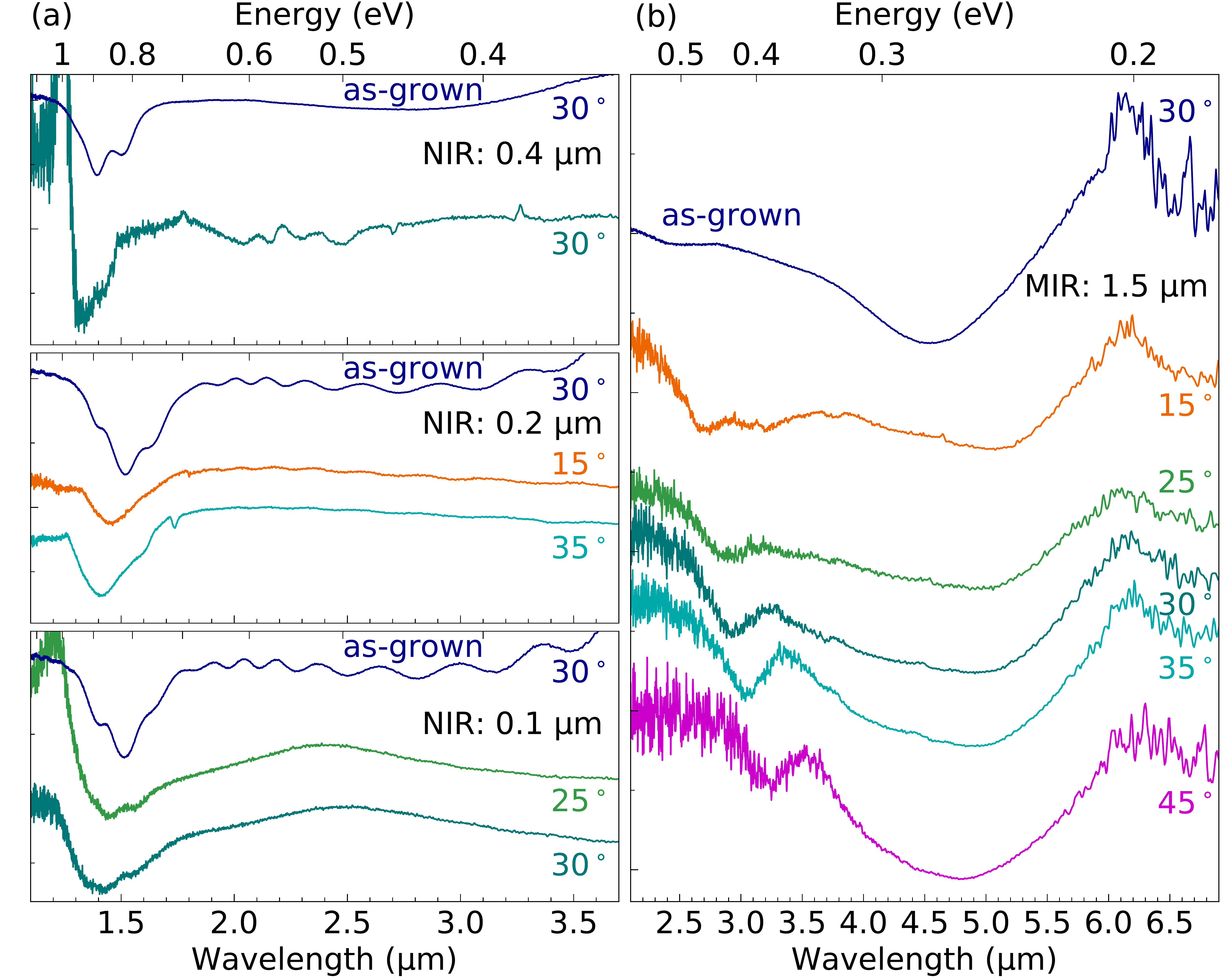}
\caption{\label{fig:ftir2}Measured FTIR transmittance of the as-grown compared with the top-down-processed samples for (a) the sample NIR and (b) the sample MIR. The diameter of the pillars for the different panels in (a) is 0.4, 0.2, and 0.1~$\upmu$m (top to bottom), while in (b) it is 1.5~$\upmu$m. The incident angle  $\theta_\mathrm{in}$ as defined in figure~\ref{fig:sem}(c) is indicated for each curve.}
\end{figure*}

And now for something completely different. Concerning the intersubband absorption, room-temperature FTIR transmittance spectra for the samples prior to (as-grown) and after top-down processing are given in figures~\ref{fig:ftir1}--\ref{fig:ftir2}. For the as-grown samples, a clear absorption dip appears at $\lambda=1.5~\upmu$m (0.83~eV) for sample NIR and at $\lambda=4.4~\upmu$m (0.28~eV) for sample MIR. These energies are in good agreement with the \textbf{k}$\cdot$\textbf{p} calculations given above. 

For the sample NIR etched down to micropillars ($d=1.5~\upmu$m), figure~\ref{fig:ftir1}(a) displays the IR transmittance measured at different angles, compared to a measurement of the same sample piece recorded before etching. After etching, the transmittance spectra exhibit several features which shift towards longer wavelenths with increasing measurement angle. Such a shift in wavelength would not be expected for intersubband absorption. As the lattice constant of the pillar array $a_\mathrm{p}$ is on the order of the investigated wavelength range, it is tempting to ascribe these features to photonic crystal stop bands \cite{Krall_2015, Pitruzzello_2018}. However, photonic crystal features are expected to blueshift when increasing the incident angle \cite{Rosenberg_1997}. 

To understand the transmittance spectra, we have carried out electrodynamics simulations accounting for the specific geometry of this sample. The simulated transmission spectra as a function of the incident angle are plotted in figure~\ref{fig:ftir1}(b). These spectra contain features, which blueshift and redshift with the incident angle (see the blue an red guides for the eye, respectively). 
Let us first focus on the redshifting features. From the electric field pattern, feature 1 can be ascribed to a guided TM mode in the remaining AlN buffer, loosely confined between the pillars and the sapphire substrate (see figure~\ref{fig:sem}), partially leaking back towards the substrate. At the same frequency and angle of incidence, the TE wave is transmitted towards the air side, mainly leaking out from the pillar facets and resulting in a peak in the TM/TE transmittance spectrum. In the case of feature 2, only the TE wave is loosely confined in the AlN template while the TM wave leaks out. Feature 3 also corresponds to a difference in the outcoupling to the air side. The TM wave is almost completely reflected towards the substrate, while the TE wave leaks out over a broad spectral range.

Turning to the blueshifting features highlighted by blue lines, which correspond to photonic crystal effects, i.e.\ modes where the electric field is mainly confined in either the pillar volume or the air around them. These modes are also resolved in the experimental spectra, though with a much smaller amplitude. An analysis of the far-field pattern in the substrate at these frequencies reveals that they emit light in directions different from the ballistic reflection. Hence, they might be only partially collected by the finite numerical aperture in the experiment. The limited resolution of the spectrometer and imperfections in the fabrication can also reduce their apparent strength.

To complement the transmittance data, figure~\ref{fig:ftir1}(c) shows the intersubband absorption spectra obtained from the simulations by integrating the resistive losses in the QWs. The simulated absorption demonstrates that the total strength of the intersubband absorption does not vary much with respect to the as-grown sample. The fact that the intersubbnand absorption is not clearly identified in the measured transmittance spectra is explained by the much higher magnitude of the refraction effects. In the end, with such a sizeable absorption, these structures could still be employed in optically structured IR detectors. 

Moving to conditions where refractive and photonic crystal effects are negligible, i.e.\ pillar diameters significantly smaller than the intersubband absorption wavelength \cite{Krall_2014}, it becomes possible to observe the intersubband absorption. Normalized IR transmittance spectra for sample NIR with sub-wavelength nanopillar diameters of $d=0.4~\upmu$m, $d=0.2~\upmu$m, and $d=0.1~\upmu$m are shown in figure~\ref{fig:ftir2}(a) (top to bottom). In all three samples, the dip related to the intersubband absorption, located around $1.5~\upmu$m, is clearly visible both before and after the etching of the nanopillars. Spectra recorded at different angles show that this feature does not shift, which confirms that it originates from the intersubband transition. The magnitude of the absorption dip at an angle of 30--35$^\circ$ (which implies $3.8\pm0.4$ waveguide passes) is $18\pm2$\% per pass in the as-grown samples and $19\pm3$\% per pass in the samples etched into nanopillars, which means that there is no degradation of the total intersubband absorption in spite of the fact that 90\% of the material was etched away, which is in good agreement with the predictions from the numerical simulations. Table~\ref{tab:samples} also shows the FWHM of the absorption before and after etching. This value is essentially unchanged except for the smallest pillar diameter ($d=0.1~\upmu$m). As with the PL intensity, we do not have a significant degradation of the intersubband absorption characteristics, even though all the measurements are performed at room temperature. However, note that the center of the absorption feature is slightly blueshifted for the nanopillars as summarized in table~\ref{tab:samples}. The blueshift amounts to 0.04--0.12~$\upmu$m (20--70~meV higher energy) without apparent trend with the pillar diameter. This shift cannot be associated to misfit relaxation, since XRD data show no variation in the strain state of the superlattice. The shift is not related either to carrier trapping, since a decrease in the density of carriers available for absorption would shift the emission towards longer wavelengths \cite{Helman_2003}.

In the case of sample MIR, figure~\ref{fig:ftir2}(b) shows the IR transmittance of measured before and after etching to $d=1.5~\upmu$m pillars. After etching, the intersubband absorption band centered around $\lambda=4.6~\upmu$m still dominates the spectra. However, an additional refraction feature is observed in the range of 2.5--3~$\upmu$m. For small angles, the central wavelength is redshifted (by 0.2~$\upmu$m) compared with the measurement for the as-grown sample, as the absorption band is broadened towards longer wavelengths, which could be due to the overlap with additional features in the transmittance spectra. Again, there is no variation of the magnitude of the absorption in spite of the reduction of the total absorbing material by 80\%.

\section{Conclusions and Outlook}

We presented a systematic analysis of intersubband absorption in group-III nitride micro- and nanopillars for different pillar diameters and absorption wavelengths in the near- and mid-IR. As a consequence of the top-down processing, a relaxation of the strain in the GaN buffer layer is observed in XRD measurements, while reciprocal space maps show that the GaN/AlN superlattice is fully relaxed already for the as-grown samples. In line with XRD results, the PL emission of the superlattice does not show any spectral shift for the processed samples. At the same time, the improved in- and outcoupling of light for the pillar arrays compensates for the etching away of a large part of the emitting material as seen from the PL intensities at 10~K. Only at room temperature, the non-radiative recombination at the pillar sidewalls has a moderate effect on the PL intensity. 

Concerning the IR absorption, we showed that when the spacing of the pillar array is comparable to the wavelength of the investigated spectral region, the IR transmittance spectra are dominated by refraction effects, as well as photonic crystal modes, and the intersubband absorption is masked by these features. However, electrodynamics simulations indicate that the intersubband absorption is still present.  For subwavelength pillar arrays, the intersubband absorption is clearly observed. We have shown that the magnitude and linewidth of the intersubband absorption is preserved in spite of the low filling factor even when 80\% or 90\% of the material is etched away. This result is in contrast to studies of micropillar quantum cascade lasers, where a high fill factor is required to achieve sufficient gain \cite{Amanti_2013}. As a perspective, this work is opening the pathway for a microstructured group-III nitride quantum-well IR photodetector and can be seen as a motivation to pursue the investigation of group-III nitride pillars also for intersubband emitter structures.

\ack
Sample processing has been carried out in the Nanofab cleanroom of CNRS Institut N\'eel and the PTA cleanroom of CEA-Grenoble. The authors would like to thank Bruno Fernandez at Nanofab for technical support with the laser lithography, as well as Luca Redaelli and Yoann Cur\'e for their assistance with first simulations. Financial support from the EU ERC-StG ``TeraGaN'' (\#278428) and from the GANEX program (ANR-11-LABX-0014) is acknowledged.

\bibliography{Top-down_NWs}

\end{document}